\documentstyle[12pt]{article}

\begin{document}

\pagenumbering{roman} 

 

\begin{center}
{\em Quantum anharmonic oscillator and its statistical properties in the first
quantization scheme}
\end{center}

\begin{center}
Maciej M. Duras
\end{center}

\begin{center}
Institute of Physics, Cracow University of Technology, 
ulica Podchor\c{a}\.zych 1, PL-30084 Cracow, Poland.
\end{center}

\begin{center}
Email: {\tt mduras @ riad.usk.pk.edu.pl}
\end{center}

\begin{center}
AD 2007 November 21st
\end{center}

\begin{center}
\end{center}

\begin{abstract}

A family of quantum anharmonic oscillators is studied in any finite spatial
dimension in the scheme of
first quantization and the investigation of their eigenenergies is presented. 
The statistical properties of the calculated eigenenergies are compared 
with the theoretical predictions inferred from the Random Matrix theory. 
Conclusions are derived.

\end{abstract}

\section{Motivation}

The quantum harmonic oscillator proved to be a fructuous model of many physical systems:
quantum electromagnetic field or systems of atoms (ions, nuclei) in ideal crystals
interacting via harmonic attractive force, \, {\sl etc.}
In the former case the excitation particles or quanta of the electromagnetic field
are called photons \cite{Einstein1905,Lewis1926}
whereas in the latter case the elementary excitation particles of vibrations
of crystal lattice or quanta of the sound field are named phonons \cite{Einstein1907,Debye1912}.
Both of these quantum fields are bosonic ones \cite{Feynman1972,FetterWalecka1971}.
Also in the case of the interaction of the quantum electromagnetic field with
the matter field via dipolar electrostatic interaction the quantum harmonic
oscillator is hugely investigated.
The harmonic potential energy is only an approximation for the real anharmonic
potential energy of mutual interaction between atoms (ions, nuclei) in real crystals.
Therefore the motivation of the present work is a more realistic description
of quantum anharmonical systems.

\section{Quantum harmonic oscillator in $D=1$ spatial dimension}

{\sl Firstly:} Our study commences to concentrate on the case of 
simple quantum harmonic oscillator in $D=1$ spatial dimension.
The dimensionless Cartesian coordinate is denoted by $x_j$ and its
conjugated dimensionless linear momentum is $p_j, j \geq 1$. 
Let us consider the Hilbert space:
\begin{equation}
{\cal V}_1 = L^2({\bf R}, {\bf C}, {\rm d}x_j),
\label{Hilbert-space-L2-D=1}
\end{equation}
of the complex-valued wave functions $\Psi$ that are modulus square integrable
on the set ${\bf R}$ of the real numbers,
and ${\bf C}$ is the set of the complex numbers.
The Hilbert space ${\cal V}_1$ is a separable space
and its orthonormal basis ${\cal B}_1$ is a set of Hermite's functions $\Psi_{r_j}$ 
(Fock's functions, eigenfunctions of the dimensionless 
Hamiltonian operator $\hat{{\cal H}}_{(j)}$ 
of the quantum harmonic oscillator in $D=1$ spatial dimension):
\begin{equation}
\Psi_{r_j}(x_j) = N_{r_j} H_{r_j}(x_j) \exp(-\frac{1}{2}x_j^2), 
N_{r_j} = [ \sqrt{\pi} r_j! 2^{r_j} ]^{-\frac{1}{2}}, r_j \in {\bf N},
\label{Hermite-function-definition}
\end{equation}
where ${\bf N}$ is a set of natural numbers including zero,
whereas:
\begin{equation}
H_{r_j}(x_j) = (-1)^{r_j} \exp(x_j^2) \frac{{\rm d}^{r_j}}{{\rm d}x_j^{r_j}} \exp(-x_j^2), 
\label{Hermite-polynomial-definition}
\end{equation}
is $r_j$th Hermite's polynomial \cite{Davydov1963}.
We assume from now that all the investigated operators are dimensionless (nondimensional).
The dimensionless quantum operator $\hat{x}_j$ of the $j$th component 
of the position vector (radius vector) operator ${\hat{\bf x}}$
acts on the  Hermite's basis $\Psi_{r_j}$ function as follows \cite{Davydov1963}:
\begin{equation}
\hat{x}_j \Psi_{r_j} = \sqrt{\frac{r_j}{2}} \Psi_{r_j-1} 
+ \sqrt{\frac{r_j+1}{2}} \Psi_{r_j+1}, 
\label{x_j-operator-Hermite-function-action}
\end{equation} 
whereas the quantum operator $\hat{p}_j$ of the $j$th component 
of linear momentum vector operator ${\hat {\bf p}}$
is given in the basis by the following formula:
\begin{equation}
\hat{p}_j \Psi_{r_j} = \frac{1}{i} \sqrt{\frac{r_j}{2}} \Psi_{r_j-1} 
- \frac{1}{i} \sqrt{\frac{r_j+1}{2}} \Psi_{r_j+1}. 
\label{p_j-operator-Hermite-function-action}
\end{equation}  
Matrix elements $(x_j)_{l_j,r_j}$ and $(p_j)_{l_j,r_j}$ 
of these operators equal correspondingly:
\begin{eqnarray}
& & (x_j)_{l_j,r_j} = \langle \Psi_{l_j} | \hat{x}_j \Psi_{r_j} \rangle_{{\cal V}_1} 
=\int_{-\infty}^{\infty} \Psi_{l_{j}}^{\star}(x_j) x_j \Psi_{r_{j}}(x_j){\rm d}x_j =
\nonumber \\
& & =\sqrt{\frac{r_j}{2}} \delta_{l_j,r_j-1} 
+ \sqrt{\frac{r_j+1}{2}} \delta_{l_j,r_j+1}, 
\label{x_j-operator-Hermite-function-matrix-element} \\
& & (p_j)_{l_j,r_j} = \langle \Psi_{l_j} | \hat{p}_j \Psi_{r_j} \rangle_{{\cal V}_1} = 
\frac{1}{i} \sqrt{\frac{r_j}{2}} \delta_{l_j,r_j-1} 
- \frac{1}{i} \sqrt{\frac{r_j+1}{2}} \delta_{l_j,r_j+1}, 
\label{p_j-operator-Hermite-function-matrix-element}
\end{eqnarray}  
where
\begin{equation}
\delta_{l_j,r_j} = 
\left\{ 
\begin{array}{cc}
1, & l_j=r_j \\
0, & l_j \neq r_j \\
\end{array}
\right. , 
\label{Kronecker-delta-definition}
\end{equation}
is discrete Kronecker's delta (it is not continuous Dirac's delta distribution $\delta_{\bf D}$).
The quantum operator $\hat{x}_j^2$ of the square of the operator $\hat{x}_j$
and quantum operator $\hat{p}_j^2$ of the square 
of the operator $\hat{p}_j$ have the following representation in the basis ${\cal B}_1$:
\begin{eqnarray}
\hat{x}_j^2 \Psi_{r_j} = & 
\frac{1}{2} 
[ \sqrt{r_j-1} \sqrt{r_j} \Psi_{r_j-2}
+ (2r_j + 1) \Psi_{r_{j}} 
+ \sqrt{r_j+1} \sqrt{r_j+2} \Psi_{r_j+2} ], 
\label{x_j^2-operator-Hermite-function-action} \\
\hat{p}_j^2 \Psi_{r_j} = &
\frac{1}{2} 
[ -\sqrt{r_j-1} \sqrt{r_j} \Psi_{r_j-2}
+ (2r_j + 1) \Psi_{r_j} 
- \sqrt{r_j+1} \sqrt{r_j+2} \Psi_{r_j+2} ], 
\label{p_j^2-operator-Hermite-function-action}
\end{eqnarray}
and their matrix elements $(x_j^2)_{l_j,r_j}, (p_j^2)_{l_j,r_j}$ read:
\begin{equation}
(x_j^2)_{l_j,r_j} = 
\frac{1}{2} 
[ \sqrt{r_j-1} \sqrt{r_j} \delta_{l_j,r_j-2}
+ (2r_j + 1) \delta_{l_j,r_j} 
+ \sqrt{r_j+1} \sqrt{r_j+2} \delta_{l_j,r_j+2} ],
\label{x_j^2-operator-Hermite-function-matrix-element} 
\end{equation}
\begin{equation}
(p_j^2)_{l_j,r_j} =
\frac{1}{2} 
[ - \sqrt{r_j-1} \sqrt{r_j} \delta_{l_j,r_j-2}
+ (2r_j + 1) \delta_{l_j,r_j} 
- \sqrt{r_j+1} \sqrt{r_j+2} \delta_{l_j,r_j+2}]. 
\label{p_j^2-operator-Hermite-function-matrix-element}
\end{equation} 
The {\sl dimensionless (nondimensional)} 
quantum Hamiltonian operator $\hat{{\cal H}}_{(j)}$ of the quantum harmonic operator
in one spatial dimension is defined as follows:
\begin{equation}
\hat{{\cal H}}_{(j)} = \hat{p}_j^2 + \hat{x}_j^2. 
\label{H_j-operator-harmonic-1D-definition}
\end{equation}  
The basis function $\Psi_{r_j}$ is its eigenfunction:
\begin{equation}
\hat{{\cal H}}_{(j)} \Psi_{r_j} = 
\epsilon_{r_j} \Psi_{r_j}, 
\label{H_j-operator-harmonic-1D-Hermite-function-action}
\end{equation}
therefore its matrix element $({\cal H}_{(j)})_{l_j,r_j}$ is equal:
\begin{equation}
({\cal H}_{(j)})_{l_j,r_j} = 
\epsilon_{r_j} \delta_{l_j,r_j}, 
\label{H_j-operator-harmonic-1D-Hermite-function-matrix-element}
\end{equation} 
where
\begin{equation}
\epsilon_{r_j} = 
2r_j + 1, 
\label{H_j-operator-harmonic-1D-Hermite-function-eigenvalue}
\end{equation} 
is the $r_j$th eigenenergy of $\hat{{\cal H}}_{(j)}$.
The eigenenergies are simply all odd natural numbers, 
and the quantum Hamiltonian is diagonal operator, 
and its matrix representation is diagonal $\infty \times \infty$ matrix.
Note, that if one introduces the notion 
of nearest neighbour energy spacing (NNS) $s_{r_j}$ between two adjacent eigenenergies:
\begin{equation}
s_{r_j} = \epsilon_{r_j+1} - \epsilon_{r_j}, 
\label{nearest-neighbour-spacing-definition}
\end{equation} 
then for the quantum harmonic oscillator it holds:
\begin{equation}
s_{r_j} = 2 = {\rm const}, 
\label{nearest-neighbour-spacing-1DIM_quantum_harmonic_oscillator}
\end{equation} 
so the eigenenergies are equidistant. 
Let us consider first $N$ consecutive energy levels 
$\epsilon_{r_j}, r_j=0, \cdots ,(N-1), N \geq 2$.
One spans over the set of the eigenvectors $\Psi_{r_j}, r_j=0, \cdots ,(N-1), N \geq 2,$ 
a Hilbert space ${\cal V}_{1,N}$ that is a subspace of the Hilbert space ${\cal V}_{1}$.
The truncated Hilbert space ${\cal V}_{1,N}$
is isomorphic to the $N$th Cartesian product ${\bf C}^N$ of complex spaces ${\bf C}$:
${\cal V}_{1,N} \equiv {\bf C}^N$.
Here $N$ is a complex dimension of
truncated Hilbert space ${\cal V}_{1,N}$ and of ${\bf C}^N$. 
The eigenfunctions of ${\bf C}^N$
are $N$-component constant complex vectors (analogs of $N$-spinors),
and the operators acting on it are $N \times N$ {\sl deterministic} complex-valued matrices.
The probability distribution $P_{N-1}$ of the spacings
is discrete one point distribution for any finite value of $N, N \geq 2$: 
\begin{equation}
P_{N-1}(s) = \frac{1}{N-1} \delta_{s,2},
\label{nearest-neighbour-spacing-distribution-finite-N}
\end{equation} 
tending in the thermodynamical limit $N \rightarrow \infty$
to the singular Dirac's delta distribution $\delta_{\bf D}$: 
\begin{equation}
P_\infty(s)=\delta_{\bf D}(s-2). 
\label{nearest-neighbour-spacing-distribution-infinite-N}
\end{equation} 

{\sl Secondly,} let us perform more difficult task 
consisting of calculating all the flip-flop transition amplitudes
(hopping amplitudes) from the quantum state $\chi_{r_j}^{s_j} = \hat{x}_j^{s_j} \Psi_{r_j}$ 
to the quantum state $\Psi_{l_j}$ ($s_j \geq 0$):
\begin{equation}
(m_{s_j})_{l_j,r_j}=(x_j^{s_j})_{l_j,r_j} = 
\langle \Psi_{l_j} | \hat{x}_j^{s_j} \Psi_{r_j} \rangle_{{\cal V}_1} 
= \int_{-\infty}^{\infty} \Psi_{l_j}^{\star}(x_j) x_j^{s_j} \Psi_{r_j}(x_j){\rm d}x_j. 
\label{x_j-operator-s_jth-transition-amplitude}
\end{equation} 
Physically, the transition amplitude $(m_{s_j})_{l_j,r_j}$ is connected
with the processes of emissions and/or absorptions of $s_j$ phonons,
because:
\begin{equation}
\hat{x}_j^{s_j} = \sqrt{2}^{s_j} (\hat{a}_j + \hat{a}_j^{+})^{s_j}, 
\label{x_j-operator-annihilation-creation-operators-relation}
\end{equation} 
where $\hat{a}_j, \hat{a}_j^{+},$ are the bosonic single phonon 
annihilation and creation operators in one spatial dimension, respectively, and:
\begin{equation}
\hat{a}_j \Psi_{r_j} = \sqrt{r_j} \Psi_{r_j-1},
\hat{a}_j^{+} \Psi_{r_j} = \sqrt{r_j+1} \Psi_{r_j+1}. 
\label{x_j-operator-annihilation-creation-operators-definition}
\end{equation} 
One can calculate the lower transition amplitudes manually, {\sl e. g.}, 
using recurrence relations, matrix algebra, {\sl etc.},
but it is tedious (even for $3 \leq s_j \leq 6$). 
If one wants to calculate {\sl all} the transition amplitudes then he must return 
to the beautiful XIX century mathematics methods
and after some reasoning he obtains the exact formula:
\begin{eqnarray}
& & (m_{s_j})_{l_j,r_j}= 
\nonumber \label{x_j-operator-s_jth-transition-amplitude-matrix-element-1} \\ 
& & = [1 - (-1)^{s_j+l_j+r_j}] \sum_{\lambda_j=0}^{[l_j/2]} \sum_{\kappa_j=0}^{[r_j/2]} 
[ (-1)^{\lambda_j+\kappa_j} 
\frac{\sqrt{l_j!}}{\lambda_j! (l_j-2\lambda_j)!} 
\cdot \frac{\sqrt{r_j!}}{\kappa_j! (r_j-2\kappa_j)!} \cdot 
\nonumber \label{x_j-operator-s_jth-transition-amplitude-matrix-element-2} \\
& & \cdot 2^{\frac{l_j}{2}+\frac{r_j}{2}-2\lambda_j-2\kappa_j-1}
\cdot \Gamma(\frac{s_j+l_j+r_j-2\lambda_j-2\kappa_j+1}{2}) ], 
\label{x_j-operator-s_jth-transition-amplitude-matrix-element-3} 
\end{eqnarray}
where $[\cdot]$ is entier (step) function, $\Gamma$ is Euler's gamma function
(compare our result Eq. (\ref{x_j-operator-s_jth-transition-amplitude-matrix-element-3}) 
with the formulae in \cite{Graffi1973,Balsa1983}).
Therefore, the matrix representations of the 
even power operators ${\hat{x}_j^{2p_j}}$ 
in the basis ${\cal B}_1$
are hermitean (symmetrical real) matrices with nonzero diagonal 
and nonzero $p_j$ subdiagonals 
(and nonzero $p_j$ superdiagonals), 
where the distance of the nearest superdiagonals (or subdiagonals) is 2 
(the diagonal is also distant by 2 from the nearest super- and sub-diagonal),
whereas the odd power operators ${\hat{x}_j^{2p_j+1}}$ in the basis ${\cal B}_1$
are hermitean (symmetrical real) matrices with zero diagonal 
and $p_j$ nonzero subdiagonals 
(and $p_j$ nonzero superdiagonals), 
where the distance of the nearest superdiagonals (or subdiagonals) is 2 
(the nearest super- and sub-diagonal are also distant by 2).
The physical interpretation of the superdiagonals (or subdiagonals) 
is connected with the absorption (or emission) of phonons.

\section{Quantum anharmonic oscillator in $D=1$ spatial dimension}

{\sl Thirdly,} we are ready to deal with the quantum anharmonic oscillator 
in $D=1$ spatial dimension.
Its {\sl dimensionless} Hamiltonian operator 
$\hat{{\cal H}}_{(j), {\rm anharm}}^{S_j}$ reads:
\begin{equation}
\hat{{\cal H}}_{(j), {\rm anharm}}^{S_j} 
=\hat{{\cal H}}_{(j)}+ \sum_{s_j=0}^{S_j} a_{s _j} \hat{x}_j^{s_j}, 
\label{H_j-operator-anharmonic-1D-S}
\end{equation}
where $S_j$ is a degree of the anharmonicity of the oscillator, 
and the prefactors $a_{s_j}$ are the strengths of anharmonicity.
The matrix elements of the anharmonic Hamiltonian operator are:
\begin{equation}
(\hat{{\cal H}}_{(j), {\rm anharm}}^{S_j})_{l_j,r_j} 
=\epsilon_{r_j} \delta_{l_j,r_j} + \sum_{s_j=0}^{S_j} a_{s_j} (x_j^{s_j})_{l_j,r_j}
=\epsilon_{r_j} \delta_{l_j,r_j} + \sum_{s_j=0}^{S_j} a_{s_j} (m_{s_j})_{l_j,r_j}, 
\label{H_j-operator-anharmonic-1D-S-matrix-element}
\end{equation}
where the representation of the quantum anharmonic oscillator 
in the quantum harmonic oscillator basis ${\cal B}_1$
is mathematically correct, because the basis ${\cal B}_1$ is a complete set, 
and the Hilbert space of the eigenfunctions
of the anharmonic oscillator is isomorphic 
to the Hilbert space ${\cal V}_1$ for the harmonic oscillator, 
provided that the total potential energy ${\cal U}_{(j), {\rm total}}^{S_j}$ 
of the anharmonic oscillator: 
\begin{equation}
{\cal U}_{(j), {\rm total}}^{S_j}(x_j) = x_j^2+ {\cal U}_{(j), {\rm anharm}}^{S_j}(x_j), 
\label{U_j-total-potential-energy-1D-S}
\end{equation}
is bounded from below (there are no scattering eigenstates),
where the anharmonic potential energy ${\cal U}_{(j), {\rm anharm}}^{S_j}$ reads:
\begin{equation}
{\cal U}_{(j), {\rm anharm}}^{S_j}(x_j) = \sum_{s_j=0}^{S_j} a_{s_j} x_j^{s_j}. 
\label{U_j-anharmonic-potential-1D-S}
\end{equation}
It suffices that the degree of the anharmonicity $S_j=2S'_j$ is an even number 
and that the strength of anharmonicity $a_{S_j}$ is strictly positive: 
$a_{S_j} > 0$, so that ${\cal U}_{(j), {\rm total}}^{S_j}(x_j) \rightarrow \infty$
for $|x_j| \rightarrow \infty$.

{\sl Fourthly,} we repeat the ``Bohigas conjecture'' 
that the fluctuations of the spectra of the quantum systems
that correspond to the chaotic systems generally obey 
the spectra of the Gaussian random matrix ensembles.
The quantum integrable systems correspond to the classical integrable systems 
in the semiclassical limit
\cite{Bohigas1984,OzoriodeAlmeida1988}.
The probability distributions $P_\beta$ of the nearest neighbour spacing 
for the Gaussian orthogonal ensemble GOE(2) of $2 \times 2$ Gaussian distributed
real-valued  symmetric random matrix variables ($\beta=1)$,
for the Gaussian unitary ensemble GUE(2) of $2 \times 2$ Gaussian distributed
complex-valued hermitean random matrix variables ($\beta=2)$,   
for the Gaussian symplectic ensemble GSE(2) of $2 \times 2$ Gaussian distributed
quaternion-valued selfdual hermitean random matrix variables ($\beta=4$),
and for the Poisson ensemble (PE) of the random diagonal matrices 
with homogeneously distributed eigenvalues
on the real axis ${\bf R}$ are given by the formulae:
\begin{equation}
P_\beta(s) = \theta(s) A_{\beta} s^{\beta} \exp(-B_{\beta} s^2), 
\label{P_beta-NNS-distribution-GOEGUEGSE} 
\end{equation} 
for the Gaussian ensembles, 
and
\begin{equation}
P_0(s) = \theta(s) \exp(- s), 
\label{P_beta-NNS-distribution-PE} 
\end{equation} 
for the Poisson ensemble,
where $\theta$ is Heaviside's unit step function   
\cite{Haake1990,Guhr1998,Mehta19900,Reichl1992,Bohigas1991,Porter1965,Brody1981,Beenakker1997,Ginibre1965,Mehta19901}. 
The constants:
\begin{equation}
A_\beta = 2 \frac{\Gamma^{\beta+1}((\beta+2)/2)}
                 {\Gamma^{\beta+2}((\beta+1)/2)}
\qquad {\rm and} \qquad 
B_\beta = \frac{\Gamma^2((\beta+2)/2)}
               {\Gamma^2((\beta+1)/2)},
\label{A_beta-B_beta-constants}
\end{equation}
are given by the formulae: $A_1=\pi/2$, $B_1=\pi/4$ (GOE),
$A_2=32/\pi^2$, $B_2=4/\pi$ (GUE), and $A_4=262144/729\pi^3$,
$B_4=64/9\pi$ (GSE), respectively.
For the Gaussian ensembles the energies are characterized by the ``level repulsion'' 
of degree $\beta$ near the origin
(near the vanishing spacing $s=0$), 
and the probability distributions $P_{\beta}$ vanish at the origin, 
and the quantum system
with ``level repulsion'' is cast to the class of quantum chaotic systems.
For the Poisson ensembles the energies are characterized 
by the ``level clustering'' near the origin, 
and the probability distributions $P_0$ has maximum at the origin, 
and the quantum system
with ``level clustering'' are treated as the quantum integrable system. 
After many numerical experiments conducted with different quantum anharmonic oscillators 
(up to the sextic quantum anharmonic oscillator $S_j=6$) 
we draw conclusion that majority of them behaves like quantum
integrable systems, the eigenenergies tend to cluster, 
the histogram of nearest neighbour spacing is closer to
the $P_0$ distribution resulting from the Poisson ensemble
\cite{Duras1996}.

\section{Quantum harmonic oscillator in $D \geq 1$ spatial dimensions}

{\sl Fifthly:} Quantum harmonic oscillator in $D$ spatial dimensions 
is a solvable analytically model.
In order to make the deliberations easier we reduce our present interest 
to the first quantization case.
Therefore the relevant Hilbert space ${\cal V}_D$ is isomorphic 
to a $D$-dimensional tensor (Cartesian) product
of the one-dimensional Hilbert spaces  ${\cal V}_1$:
\begin{equation}
{\cal V}_D \equiv \bigotimes_{j=1}^{D} {\cal V}_1,
\label{Hilbert-space-L2-D-tensor-product}
\end{equation}
whereas the ${\bf r}$th harmonic oscillator's eigenfunction $\Psi_{\bf r}$ 
in $D$ dimensions is, neither symmetrized nor antisymmetrized, 
tensor product of the eigenfunctions in one dimension:
\begin{equation}
\Psi_{\bf r}({\bf x}) = \prod_{j=1}^{D} \Psi_{r_{j}}(x_j), {\bf x}=(x_1,...,x_D) \in {\bf R}^D, 
{\bf r}=(r_1,...,r_D) \in {\bf N}^D,
\label{eigenfunction-tensor-product-definition-D}
\end{equation}
where we used boldface font for the $D$-tuples ${\bf x},$ and ${\bf r}.$
It follows that:
\begin{equation}
\Psi_{\bf r}({\bf x}) = N_{\bf r} H_{\bf r}({\bf x}) \exp(-\frac{1}{2}{\bf x}^2), 
N_{\bf r} = \prod_{j=1}^{D} N_{r_{j}},  
H_{\bf r}({\bf x})= \prod_{j=1}^{D} H_{r_{j}}(x_j),
\label{eigenfunction-Hermite-function-definition-D}
\end{equation}
whereas 
\begin{equation}
{\bf x}^2={\bf x} \cdot {\bf x}= \sum_{j=1}^{D}(x_j)^2.
\label{D-tuple-scalar-product}
\end{equation}
One can also draw a conclusion that the Hilbert space:
\begin{equation}
{\cal V}_D = L^2({\bf R}^D, {\bf C}, {\rm d}x), D \geq 1,
\label{Hilbert-space-L2-D}
\end{equation}
is composed of the complex-valued wave functions $\Psi$ that are modulus square integrable
on the set ${\bf R}^D$.
The Hilbert space ${\cal V}_D$ is separable space,
and its orthonormal basis ${\cal B}_D$ is a set of Hermite's functions $\Psi_{\bf r}$ in $D$ dimensions 
(Fock's functions in $D$ dimensions).
The {\sl dimensionless (nondimensional)} 
quantum Hamiltonian operator $\hat{{\cal H}}_D$ of the quantum harmonic oscillator in $D$ dimensions 
is a sum of quantum Hamiltonian operators $\hat{{\cal H}}_{(j)}$ 
of the quantum harmonic oscillators in one dimension:
\begin{equation}
\hat{{\cal H}}_D = \sum_{j=1}^{D} \hat{{\cal H}}_{(j)} 
= \sum_{j=1}^{D} (\hat{p}_j^2+ \hat{x}_{j}^2) = \hat{{\bf p}}^2 + \hat{{\bf x}}^2, 
\label{H-operator-harmonic-D-definition}
\end{equation}  
The quantum Hamiltonian operator $\hat{{\cal H}}_D$ is diagonal in Fock's basis of its
eigenfunctions $\Psi_{\bf r}$:
\begin{equation}
\hat{{\cal H}}_D \Psi_{\bf r} = 
\epsilon_{\bf r} \Psi_{{\bf r}}, 
\label{H-operator-harmonic-D-Hermite-function-action}
\end{equation}
and its matrix element $({\cal H}_D)_{{\bf l}, {\bf r}}$ is equal:
\begin{equation}
({\cal H}_D)_{{\bf l}, {\bf r}} = 
\epsilon_{\bf r} \delta_{{\bf l},{\bf r}}, 
\label{H-operator-harmonic-D-Hermite-function-matrix-element}
\end{equation} 
whereas
\begin{equation}
\epsilon_{\bf r}  
= \sum_{j=1}^{D} \epsilon_{r_{j}} = \sum_{j=1}^{D} (2 r_{j}+1), 
\label{H-operator-harmonic-D-Hermite-function-eigenvalue}
\end{equation} 
is the ${\bf r}$th eigenenergy of $\hat{{\cal H}}_D$ 
and where
\begin{equation}
\delta_{{\bf l},{\bf r}} = \prod_{j=1}^{D} \delta_{l_j,r_j}, 
\label{Kronecker-delta-definition-D}
\end{equation}
is discrete Kronecker's delta in $D$ dimensions 
(it is not continuous Dirac's delta in $D$ dimensions).
The eigenenergies $\epsilon_{\bf r}$ are simply the sums of all odd natural numbers, 
and the quantum Hamiltonian $\hat{{\cal H}}_D$ 
is (direct) sum of diagonal operators in one dimension, 
and its matrix representation is direct sum of diagonal $\infty \times \infty$ matrices
(it is poly-index matrix).
There is no unique straightforward analog of the nearest neighbour spacing $s_{r_j}$ in $D$ dimensions, 
for $D \geq 2$.

{\sl Sixthly,} let us perform very difficult task 
consisting of calculating all the flip-flop transition amplitudes (hopping amplitudes) 
from the quantum state $\chi_{\bf r}^{\bf s} = \hat{{\bf x}}^{\bf s} \Psi_{\bf r}$ 
to the quantum state $\Psi_{\bf l}$ (${\bf s} \geq {\bf 0}$):
\begin{equation}
(m_{\bf s})_{{\bf l}, {\bf r}}=({\bf x}^{\bf s})_{{\bf l}, {\bf r}} 
= \langle \Psi_{\bf l} | \hat{{\bf x}}^{\bf s} \Psi_{\bf r} \rangle_{{\cal V}_D} 
= \prod_{j=1}^{D}\int_{-\infty}^{\infty} \Psi_{l_{j}}^{\star}(x_j) x_j^{s_{j}}
\Psi_{r_{j}}(x_j){\rm d}x_j.
\label{x-operator-sth-transition-amplitude-D}
\end{equation} 
The flip-flop transition amplitude $(m_{\bf s})_{{\bf l}, {\bf r}}$ is connected with
the processes of emissions and/or absorptions of ${\bf s}$ phonons in $D$ spatial dimensions,
because:
\begin{equation}
\hat{{\bf x}}^{\bf s} = \prod_{j=1}^{D} (x_j^{s_{j}})
=\prod_{j=1}^{D} [\sqrt{2}^{s_j} (\hat{a}_j + \hat{a}_j^{+})^{s_j})]
=[\sqrt{2} (\hat{{\bf a}} + \hat{{\bf a}}^{+})]^{\bf s}, 
\label{x-operator-annihilation-creation-operators-relation-D}
\end{equation} 
where $\hat{{\bf a}}=(\hat{a}_1,...,\hat{a}_D), 
\hat{{\bf a}}^{+}=(\hat{a}_1^{+},...,\hat{a}_D^{+}),$
are the bosonic multiphonon ($D$-phonon) 
annihilation and creation operators in $D$ spatial dimensions, respectively, and:
\begin{equation}
\hat{a}_j \Psi_{\bf r} = \sqrt{r_j} \Psi_{(r_1,...,r_j-1,...,r_D)},
\hat{a}_j^{+} \Psi_{\bf r} = \sqrt{r_j+1} \Psi_{(r_1,...,r_j+1,...,r_D)}. 
\label{x-operator-annihilation-creation-operators-definition-D}
\end{equation} 
It can be easily proven that:
\begin{equation}
({\bf x}^{\bf s})_{{\bf l}, {\bf r}} 
= \prod_{j=1}^{D} (x_j^{s_{j}})_{l_j, r_j} . 
\label{x-operator-sth-power-definition-D}
\end{equation} 
One can calculate the lower transition amplitudes manually, {\sl e. g.}, 
using recurrence relations, matrix algebra, {\sl etc.},
but it is tedious (even for $3 \leq s_j \leq 6$). 
The exact formula for {\sl all} the transition amplitudes reads:
\begin{equation}
(m_{\bf s})_{{\bf l}, {\bf r}}=\prod_{j=1}^{D} (m_{s_{j}})_{l_j, r_j}.
\label{x-operator-sth-transition-amplitude-matrix-element-D} 
\end{equation}

\section{Quantum anharmonic oscillator in $D \geq 1$ spatial dimensions}

{\sl Seventhly,} we are ready to investigate the quantum anharmonic oscillator 
in $D$ spatial dimensions.
Its {\sl dimensionless} Hamiltonian operator 
$\hat{{\cal H}}_{D, {\rm anharm}}^{{\bf S}}$ reads:
\begin{equation}
\hat{{\cal H}}_{D, {\rm anharm}}^{{\bf S}} 
=\hat{{\cal H}}_D+ \sum_{{\bf s}={\bf 0}}^{{\bf S}} a_{\bf s} \hat{{\bf x}}^{\bf s}
=\hat{{\cal H}}_D+ \sum_{(s_1,...,s_D)=(0,...,0)}^{(S_1,...,S_D)} [a_{(s_1,..,s_D)} \cdot 
\prod_{j=1}^{D}(\hat{x}_j)^{s_j}], 
\label{H-operator-anharmonic-D-S}
\end{equation}
where ${\bf S}$ is a $D$-tuple of degrees of the anharmonicity of the oscillator, 
and the prefactors $a_{\bf s}$ are the strengths of anharmonicity.
The matrix elements of the anharmonic Hamiltonian operator are:
\begin{equation}
({\cal H}_{D, {\rm anharm}}^{{\bf S}})_{{\bf l}, {\bf r}} 
=\epsilon_{\bf r} \delta_{{\bf l}, {\bf r}} 
+ \sum_{{\bf s}={\bf 0}}^{{\bf S}} a_{\bf s} ({\bf{x}}^{\bf s})_{{\bf l}, {\bf r}}
=\epsilon_{\bf r} \delta_{{\bf l}, {\bf r}} 
+ \sum_{{\bf s}={\bf 0}}^{{\bf S}} a_{\bf s} ({\bf{m}}_{\bf s})_{{\bf l}, {\bf r}}, 
\label{H-operator-anharmonic-D-S-matrix-element}
\end{equation}
where the representation of the $D$-dimensional quantum anharmonic oscillator 
in the quantum harmonic oscillator basis ${\cal B}_D$
is mathematically correct, because the basis ${\cal B}_D$ is a complete set, 
and the Hilbert space of the eigenfunctions
of the anharmonic oscillator is isomorphic 
to the Hilbert space ${\cal V}_D$ for the harmonic oscillator, 
provided that
the total potential energy ${\cal U}_{D, {\rm total}}^{{\bf S}}$ 
of the quantum anharmonic oscillator in $D$ dimensions: 
\begin{equation}
{\cal U}_{D, {\rm total}}^{{\bf S}}({\bf x}) 
= {\bf x}^2+ {\cal U}_{D, {\rm anharm}}^{{\bf S}}({\bf x}), 
\label{U-total-potential-energy-D-S}
\end{equation}
is bounded from below (there are no scattering eigenstates in $D$ dimensions),
where the anharmonic potential energy ${\cal U}_{D, {\rm anharm}}^{{\bf S}}$ is:
\begin{equation}
{\cal U}_{D, {\rm anharm}}^{{\bf S}}({\bf x}) = \sum_{{\bf s}={\bf 0}}^{{\bf S}} a_{\bf s} {\bf x}^{\bf s}
=\sum_{(s_1,...,s_D)=(0,...,0)}^{(S_1,...,S_D)} [a_{(s_1,..,s_D)} \cdot 
\prod_{j=1}^{D}(x_j)^{s_j}]. 
\label{U-anharmonic-potential-D-S}
\end{equation}
It suffices that the $D$-tuple of degrees of the anharmonicity of the oscillator 
${\bf S}=2 {\bf S'}$ is composed of even numbers 
and that the strength of anharmonicity $a_{\bf S}$ is strictly positive: 
$a_{\bf S} > 0$, so that ${\cal U}_{D, {\rm total}}^{{\bf S}}({\bf x}) \rightarrow \infty$
for $|{\bf x}| \rightarrow \infty$.

{\sl Eighthly,} we repeat that the ``Bohigas conjecture'' also holds 
for the quantum oscillators in $D$ dimensions. 
Having conducted many numerical experiments with different quantum anharmonic oscillators 
(up to the sextic $(D=3)$-dimensional quantum anharmonic oscillators: $S_j=6$) 
we draw conclusion that some of them behave like quantum
integrable systems, the eigenenergies tend to cluster, 
the histograms of nearest neighbour spacing are closer to
the $P_0$ distribution resulting from the Poisson ensemble,
whereas other ones look like quantum chaotic systems, 
their eigenenergies are subject to repulsion,
the histograms of NNS are closer 
to the distributions $P_1, P_2, P_4,$ derived from the Gaussian
Random Matrix ensembles \cite{Duras1996}.



\begin{thebibliography}{12}

\bibitem{Einstein1905} 
A. Einstein, 
Annalen der Physik (Leipzig) {\bf 17} 132 (1905). 
\bibitem{Lewis1926} 
G. N. Lewis, 
Nature {\bf 118} 874 (1926). 
\bibitem{Einstein1907} 
A. Einstein, 
Annalen der Physik (Leipzig) {\bf 22} 180 (1907). 
\bibitem{Debye1912} 
P. Debye, 
Annalen der Physik (Leipzig) {\bf 39} 789 (1912). 
\bibitem{Feynman1972} 
R. P. Feynman, 
{\sl  Statistical Mechanics: A Set of Lectures} 
(W. A. Benjamin, Reading, Massachusetts, 1972).
\bibitem{FetterWalecka1971} 
A. L. Fetter, J. D. Walecka, 
{\sl Quantum theory of Many-Particle Systems} 
(McGraw-Hill Book Company, San Francisco, 1971).
\bibitem{Davydov1963}
A. S. Davydov, 
{\sl Quantum Mechanics} 
(GIFML Editors, Moscow, 1963).
\bibitem{Graffi1973} 
S. Graffi, V. Grecchi, 
Phys. Rev. D {\bf 8} 3487 (1973). 
\bibitem{Balsa1983} 
R. Balsa, M. Plo, J. G. Esteve, A. F. Pacheco, 
Phys. Rev. D {\bf 28} 1945 (1983). 
\bibitem{Bohigas1984} 
O. Bohigas, M. J. Giannoni, C. Schmidt, 
Phys. Rev. Lett. {\bf 52} 1 (1984). 
\bibitem{OzoriodeAlmeida1988} 
A. M. Ozorio de Almeida, 
{\sl Hamiltonian systems: chaos and quantization} 
(Cambridge University Press, Cambridge, 1988).
\bibitem{Haake1990}
 F. Haake, 
{\sl Quantum Signatures of Chaos} 
 (Springer-Verlag, Berlin, Heidelberg, New York, 1990), 
 Chapters 1, 3, 4, 8, pp. ~1--11, 33--77, 202--213.
\bibitem{Guhr1998} 
 T. Guhr, A. M\"uller-Groeling, H. A. Weidenm\"uller, 
 Phys. Rep. {\bf 299} 189 (1998).  
\bibitem{Mehta19900}
 M. L. Mehta, 
 {\sl Random matrices} 
 (Academic Press, Boston, 1990), Chapters 1, 2, 9, pp. ~1--54, 182--193.
\bibitem{Reichl1992}
 L. E. Reichl, 
 {\sl The Transition to Chaos In Conservative Classical 
 Systems: Quantum Manifestations} (Springer-Verlag, New York, 1992), 
 Chapter 6, pp. ~248--286.
\bibitem{Bohigas1991}
 O. Bohigas, in {\sl Proceedings of the Les Houches Summer School on Chaos and
 Quantum Physics, Session LII, 1 - 31 August 1989 
 [LES HOUCHES \'ECOLE D'\'ET\'E DE PHYSIQUE TH\'EORIQUE, 
 SESSION LII, 1 - 31 AO\^UT 1989]} 
 edited by M. - J. Giannoni, A. Voros, J. Zinn-Justin 
 (North-Holland, Amsterdam, 1991), pp. ~87--199.
\bibitem{Porter1965}
 C. E. Porter, 
 {\sl Statistical Theories of Spectra: Fluctuations} 
 (Academic Press, New York, 1965).
\bibitem{Brody1981}
 T. A. Brody, J. Flores, J. B. French, P. A. Mello, A. Pandey, S. S. M. Wong, 
 Rev. Mod. Phys. {\bf 53} 385 (1981).
\bibitem{Beenakker1997}
 C. W. J. Beenakker, 
 Rev. Mod. Phys. {\bf 69} 731 (1997).  
\bibitem{Ginibre1965} 
 J. Ginibre,  
 J. Math. Phys. {\bf 6} 440 (1965). 
\bibitem{Mehta19901}
 M. L. Mehta, 
 {\sl Random matrices} 
 (Academic Press, Boston, 1990), Chapter 15, pp. ~294--310.
\bibitem{Duras1996}
 M. M. Duras, unpublished. 

\end{thebibliography}
\end{document}